\documentclass[longauth]{aa}
\usepackage{graphicx}
\usepackage{txfonts}
\usepackage{xcolor}
\usepackage{hyperref}
\hypersetup{colorlinks, linkcolor=blue, citecolor=blue, urlcolor=blue}

\newcommand{\pivec}{\mbox{\boldmath $\pi$}}
\newcommand{\muvec}{\mbox{\boldmath $\mu$}}

\newcommand{\te}{t_{\rm E}}
\newcommand{\thetae}{\theta_{\rm E}}

\newcommand{\pie}{\pi_{\rm E}}

\newcommand{\dl}{D_{\rm L}}

\definecolor{brown}{rgb}{0.59, 0.29, 0.0}
\definecolor{darkgreen}{rgb}{0.0, 0.42, 0.24}
\definecolor{darkblue}{rgb}{0.01, 0.31, 0.59}
\definecolor{darkblue}{rgb}{0.0, 0.25, 0.42}
\definecolor{blue}{rgb}{0.0,0.0,1.0}
\definecolor{green}{rgb}{0.0,1.0,0.0}

\def\eqalign#1{\null\,\vcenter{\openup\jot
        \ialign{\strut\hfil$\displaystyle{##}$&$
        \displaystyle{{}##}$\hfil \crcr#1\crcr}}\,}

\begin{document}

\title{KMT-2021-BLG-1077L: The fifth confirmed multiplanetary system detected by microlensing}

\author{
     Cheongho~Han\inst{01} 
\and Andrew~Gould\inst{02,03}      
\and Ian A. Bond\inst{04}
\and Youn~Kil~Jung\inst{06} 
\\
(Leading authors)\\
     Michael~D.~Albrow\inst{06}   
\and Sun-Ju~Chung\inst{05}     
\and Kyu-Ha~Hwang\inst{05} 
\and Yoon-Hyun~Ryu\inst{05} 
\and In-Gu~Shin\inst{05} 
\and Yossi~Shvartzvald\inst{07}   
\and Jennifer~C.~Yee\inst{08}   
\and Weicheng~Zang\inst{09}     
\and Sang-Mok~Cha\inst{06,10} 
\and Dong-Jin~Kim\inst{06} 
\and Seung-Lee~Kim\inst{06} 
\and Chung-Uk~Lee\inst{06} 
\and Dong-Joo~Lee\inst{06} 
\and Yongseok~Lee\inst{06,11} 
\and Byeong-Gon~Park\inst{06,11} 
\and Richard~W.~Pogge\inst{03}
\and Doeon~Kim\inst{01}
\\
(The KMTNet collaboration)\\
     Fumio~Abe\inst{12}
\and Richard~K.~Barry\inst{13} 
\and David~P.~Bennett\inst{13,14} 
\and Aparna~Bhattacharya\inst{13,14}
\and Hirosane~Fujii\inst{12}
\and Akihiko~Fukui\inst{15,16} 
\and Yuki~Hirao\inst{13,17,18}
\and Yoshitaka~Itow\inst{12}
\and Rintaro~Kirikawa\inst{18} 
\and Naoki~Koshimoto\inst{18} 
\and Iona~Kondo\inst{18} 
\and Yutaka~Matsubara\inst{18}
\and Sho~Matsumoto\inst{18}
\and Shota~Miyazaki\inst{18} 
\and Yasushi~Muraki\inst{18} 
\and Greg~Olmschenk\inst{19}
\and Arisa~Okamura\inst{18} 
\and Cl\'ement~Ranc\inst{20} 
\and Nicholas~J.~Rattenbury\inst{21} 
\and Yuki~Satoh\inst{18}
\and Stela~Ishitani Silva\inst{22}
\and Takahiro~Sumi\inst{18}
\and Daisuke~Suzuki\inst{23}
\and Taiga~Toda\inst{18}
\and Paul~J.~Tristram\inst{24} 
\and Aikaterini~Vandorou\inst{19} 
\and Hibiki~Yama\inst{18}
\\
(The MOA Collaboration)\\
}

\institute{
     Department of Physics, Chungbuk National University, Cheongju 28644, Republic of Korea  \\ \email{cheongho@astroph.chungbuk.ac.kr}                  
\and Max Planck Institute for Astronomy, K\"onigstuhl 17, D-69117 Heidelberg, Germany                                                                    
\and Department of Astronomy, The Ohio State University, 140 W. 18th Ave., Columbus, OH 43210, USA                                                       
\and Institute of Information and Mathematical Sciences, Massey University, Private Bag 102-904, North Shore Mail Centre, Auckland, New Zealand          
\and Korea Astronomy and Space Science Institute, Daejon 34055, Republic of Korea                                                                        
\and University of Canterbury, Department of Physics and Astronomy, Private Bag 4800, Christchurch 8020, New Zealand                                     
\and Department of Particle Physics and Astrophysics, Weizmann Institute of Science, Rehovot 76100, Israel                                               
\and Center for Astrophysics $|$ Harvard \& Smithsonian 60 Garden St., Cambridge, MA 02138, USA                                                          
\and Department of Astronomy and Tsinghua Centre for Astrophysics, Tsinghua University, Beijing 100084, China                                            
\and School of Space Research, Kyung Hee University, Yongin, Kyeonggi 17104, Republic of Korea                                                           
\and Korea University of Science and Technology, 217 Gajeong-ro, Yuseong-gu, Daejeon, 34113, Republic of Korea                                           
\and Institute for Space-Earth Environmental Research, Nagoya University, Nagoya 464-8601, Japan                                                         
\and Laboratory for Exoplanets and Stellar Astrophysics, NASA / Goddard Space Flight Center, Greenbelt, MD 20771, USA                                    
\and Department of Astronomy, University of Maryland, College Park, MD 20742, USA                                                                        
\and Department of Earth and Planetary Science, Graduate School of Science, The University of Tokyo, 7-3-1 Hongo, Bunkyo-ku, Tokyo                       
\and Instituto de Astrof\'isica de Canarias, V\'ia L\'actea s/n, E-38205 La Laguna, Tenerife, Spain                                                      
\and Astrophysics Science Division, NASA/Goddard Space Flight Center, Greenbelt, MD20771, USA                                                            
\and Department of Earth and Space Science, Graduate School of Science, Osaka University, 1-1 Machikaneyama, Toyonaka, Osaka 560-0043, Japan             
\and Code 667, NASA Goddard Space Flight Center, Greenbelt, MD 20771, USA                                                                                
\and Zentrum f{\"u}r Astronomie der Universit{\"a}t Heidelberg, Astronomisches Rechen-Institut, M{\"o}nchhofstr.\ 12-14, 69120 Heidelberg, Germany       
\and Department of Physics, University of Auckland, Private Bag 92019, Auckland, New Zealand                                                             
\and Department of Physics, The Catholic University of America, Washington, DC 20064, USA                                                                
\and Institute of Space and Astronautical Science, Japan Aerospace Exploration Agency, 3-1-1 Yoshinodai, Chuo, Sagamihara, Kanagawa, 252-5210, Japan     
\and University of Canterbury Mt. John Observatory, P.O. Box 56, Lake Tekapo 8770, New Zealand                                                           
}
\date{Received ; accepted}

\abstract
{}
{
The high-magnification microlensing event KMT-2021-BLG-1077 exhibits a subtle and complex 
anomaly pattern in the region around the peak. We analyze the lensing light curve of the 
event with the aim of revealing the nature of the anomaly.
}
{
We test various models in combination with several interpretations: that the lens is a binary 
(2L1S), the source is a binary (1L2S), both the lens and source are binaries (2L2S), or the 
lens is a triple system (3L1S).  We search for the best-fit models under the individual 
interpretations of the lens and source systems.  
}
{
We find that the anomaly cannot be explained by the usual three-body (2L1S and 1L2S) models.
The 2L2S model improves the fit compared to the three-body models, but it still leaves 
noticeable residuals.  On the other hand, the 3L1S interpretation yields a model explaining 
all the major anomalous features in the lensing light curve.  According to the 3L1S interpretation, 
the estimated mass ratios of the lens companions to the primary are $\sim 1.56 \times 10^{-3}$ 
and $\sim 1.75 \times 10^{-3}$, which correspond to $\sim 1.6$ and $\sim 1.8$ times the 
Jupiter/Sun mass ratio, respectively, and therefore the lens is a multiplanetary system containing 
two giant planets.  With the constraints of the event time-scale and angular Einstein radius, 
it is found that the host of the lens system is a low-mass star of mid-to-late M spectral type 
with a mass of $M_{\rm h} = 0.14^{+0.19}_{-0.07}~M_\odot$, and it hosts two gas giant planets with 
masses of $M_{\rm p_1}=0.22^{+0.31}_{-0.12}~M_{\rm J}$ and $M_{\rm p_2}=0.25^{+0.35}_{-0.13}~M_{\rm J}$.  
The planets lie beyond the snow line of the host with projected separations of $a_{\perp, 
{\rm p}_1}=1.26^{+1.41}_{-1.08}~{\rm AU}$ and $a_{\perp, {\rm p}_2}=0.93^{+1.05}_{-0.80}~{\rm AU}$.  
The planetary system resides in the Galactic bulge at a distance of $\dl=8.24^{+1.02}_{-1.16}~{\rm kpc}$.  
The lens of the event is the fifth confirmed multiplanetary system detected by microlensing 
following OGLE-2006-BLG-109L, OGLE-2012-BLG-0026L, OGLE-2018-BLG-1011L, and OGLE-2019-BLG-0468L.
}
{}

\keywords{gravitational microlensing -- planets and satellites: detection}

\maketitle

\section{Introduction}\label{sec:one}
        
There are various advantages to using the microlensing method of planet detection, making it an important
complement to other planet-detection methods for the demographic study of extrasolar planets.  
These advantages include the high sensitivity to cold planets lying near and beyond the snow 
line (e.g., OGLE-2005-BLG-390Lb \citep{Beaulieu2006} and OGLE-2005-BLG-169L \citep{Gould2006}), 
the sensitivity to planets with low masses down to below Earth mass (e.g., OGLE-2016-BLG-1195Lb 
\citep{Shvartzvald2017} and KMT-2020-BLG-0414Lb \citep{Zang2021}), the unique sensitivity to 
free-floating planets that are not gravitationally bound to hosts (e.g., OGLE-2016-BLG-1540L 
\citep{Mroz2018} and KMT-2017-BLG-2820L \citep{Ryu2021}), and the sensitivity to planets with 
various types of hosts including not only regular stars but also stellar remnants (e.g., 
MOA-2010-BLG-477Lb with a white-dwarf host \citep{Blackman2021}). For a detailed and comprehensive 
discussion about the various advantages of the microlensing method, see the review paper of 
\citet{Gaudi2012}.

Microlensing is also important because of its sensitivity to planetary systems with multiple 
planets. The microlensing detection of a multiplanetary system is possible because the 
individual planets of a system induce their own caustics and the multiple planetary 
signatures can be detected if the source passes through the anomaly regions induced by the 
individual planets \citep{Han2001}. The efficiency to multiple planets is especially high 
for high-magnification events, in which the caustics induced by the individual planets lie 
in the common central magnification region around the planet host and the source passes 
through this region \citep{Gaudi1998b, Han2005}.

According to the core-accretion model of planet formation \citep{Ida2010}, multiple giant 
planets can form near and beyond the snow line, where solid grains are abundant for accretion 
into planetesimals and eventually planets. From a microlensing simulation conducted within the framework of the 
core-accretion model, \citet{Zhu2014} predicted that about 5.5\% of planetary events that 
are detectable by high-cadence microlensing surveys would exhibit signatures of multiple 
planets.

The first multiplanetary system found by microlensing is OGLE-2006-BLG-109L, which contains 
two planets with masses of $\sim 0.71~M_{\rm J}$ and $\sim 0.27~M_{\rm J}$ and semi-major 
axes of $\sim 2.3$~AU and $\sim 4.6$~AU orbiting a primary star with a mass of $\sim 0.50~M_\odot$. 
Thus, the system resembles a scaled version of our Solar System in terms of mass ratio and 
separation ratio, and the equilibrium temperatures of the planets are similar to those of Jupiter 
and Saturn \citep{Gaudi2008, Bennett2010}.  OGLE-2012-BLG-0026L is the 
second multiplanetary system, for which a G-type main sequence host star with a mass of 
$\sim 0.82~M_\odot$ contains two planets with masses of $\sim 0.11~M_{\rm J}$ and 
$\sim 0.68~M_{\rm J}$ \citep{Han2013, Beaulieu2016, Madsen2019}.  The third microlensing 
multiplanetary system is OGLE-2018-BLG-1011L, which has two identified planets with masses of 
$1.8_{-1.1}^{+3.4}~M_{\rm J}$ and $2.8_{-1.7}^{+5.1}~M_{\rm J}$ around a low-mass host star 
with a mass of $0.18_{-0.10}^{+0.33}~M_\odot$ \citep{Han2019}. The system found most recently 
is OGLE-2019-BLG-0468L \citep{Han2022b}, in which two planets with masses of $\sim 3.4~M_{\rm J}$ 
and $\sim 10.2~M_{\rm J}$ orbit a G-type host star with a mass of $\sim 0.9~M_\odot$.  We 
note that planetary signals of all these systems were detected through the high-magnification 
channel. Besides these systems, there are three candidates multiplanetary systems: 
KMT-2021-BLG-0240L \citep{Han2022a}, OGLE-2014-BLG-1722 \citep{Suzuki2018}, and KMT-2019-BLG-1953 
\citep{Han2020}. Compared to the four confirmed systems, the existence of multiple planets is 
less secure for these systems either because of the weak signals of the second planet or degeneracies 
with other interpretations of the signals.

In this work, we report the fifth confirmed multiplanetary system found by microlensing.  
The signatures of the multiple planets were found from the analysis of a high-magnification 
lensing event observed during the 2021 bulge season by two high-cadence lensing surveys, the
Korea Microlensing Telescope Network \citep[KMTNet:][]{Kim2016} and the Microlensing 
Observations in Astrophysics \citep[MOA:][]{Bond2001}.  The dense and continuous 
coverage with the use of the globally distributed multiple telescopes of the surveys 
captured the detailed structure of the short-duration anomaly, leading to detections 
of two very low-mass companions of the lens system.

The analysis leading to the discovery of the planetary system is presented as follows.  In 
Sect.~\ref{sec:two}, we mention the photometric data of the lensing event used in the analysis 
and the procedure used for data reduction.  In Sect.~\ref{sec:three}, we describe models conducted 
under various interpretations of the lensing system and explain the
modeling procedure  in detail.  In Sect.~\ref{sec:four}, we specify the type of the source star and estimate the 
angular Einstein radius. In Sect.~\ref{sec:five}, we estimate the physical parameters of the 
planetary system by conducting a Bayesian analysis of the lensing event. In Sect.~\ref{sec:six}, 
we summarize the results from our analysis and present our conclusions.

\section{Observations and data}\label{sec:two}

The multiplanetary system was found from the analysis of the microlensing event KMT-2021-BLG-1077. 
The source of the event lies in the Galactic bulge field with equatorial coordinates (RA, 
DEC)$_{\rm J2000}=$(17:45:57.39, -33:50:34.12), which correspond to the Galactic coordinates 
$(l, b)=(-4^\circ\hskip-2pt .154, -2^\circ\hskip-2pt .613)$.  The baseline brightness of the 
source before lensing magnification was $I_{\rm base}=18.73$ based on the calibrated OGLE-III 
catalog \citep{Szymanski2011}.

\begin{figure}[t]
\includegraphics[width=\columnwidth]{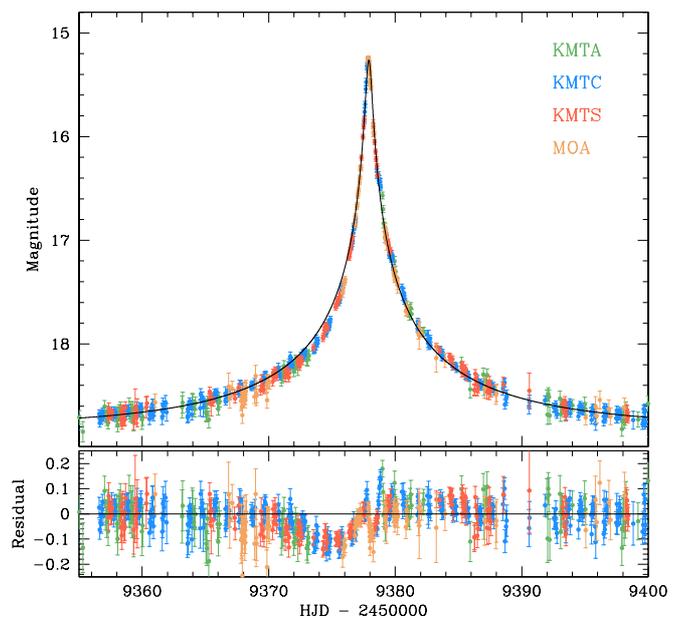}
\caption{
Light curve of the lensing event KMT-2021-BLG-1077.  The curve plotted over the data points is 
a single-lens single-source (1L1S) model obtained with the exclusion of the data lying in the 
anomaly region.  The bottom panel shows the residual from the 1L1S model.  The colors of the data 
points are set to match those of the data sets marked in the legend. 
}
\label{fig:one}
\end{figure}

The lensing event was first discovered by the KMTNet survey on 2021 June 1 (${\rm HJD}^\prime 
\equiv {\rm HJD}-2450000 = 9366.56$) using the KMT Alert Finder system \citep{Kim2018}, when 
the source became  $\sim 0.22$~mag brighter than the baseline. The KMTNet survey uses 
three identical telescopes each of which has a 1.6~m aperture and is mounted with a camera 
yielding a $2\times 2~{\rm deg}^2$ field of view.  For continuous coverage of lensing events, 
the telescopes are globally distributed in three continents of the Southern Hemisphere: the 
Siding Spring Observatory in Australia (KMTA), Cerro Tololo Inter-American Observatory in Chile 
(KMTC), and South African Astronomical Observatory in South Africa (KMTS). The event was found 
independently by the MOA survey group, who designated the event MOA-2021-BLG-173 on 2021 
June 10 (${\rm HJD}^\prime=9376.04$).  The MOA telescope, located at the Mt.~John Observatory 
in New Zealand, has a 1.8~m aperture, and it is mounted with a camera yielding a 2.2~deg$^2$ 
field of view.  In accordance with the nomenclature convention of the microlensing community 
using the event ID of the first discovery survey, we hereafter designate the event 
KMT-2021-BLG-1077.  The event reached a high magnification of $A_{\rm peak}\sim 110$ at the 
peak on 2021 June 12, and then gradually declined to the baseline.  Although there was an alert to the event well before the peak,  no follow-up observations were 
conducted to the best of our
knowledge.

\begin{figure*}[t]
\centering
\includegraphics[width=12.0cm]{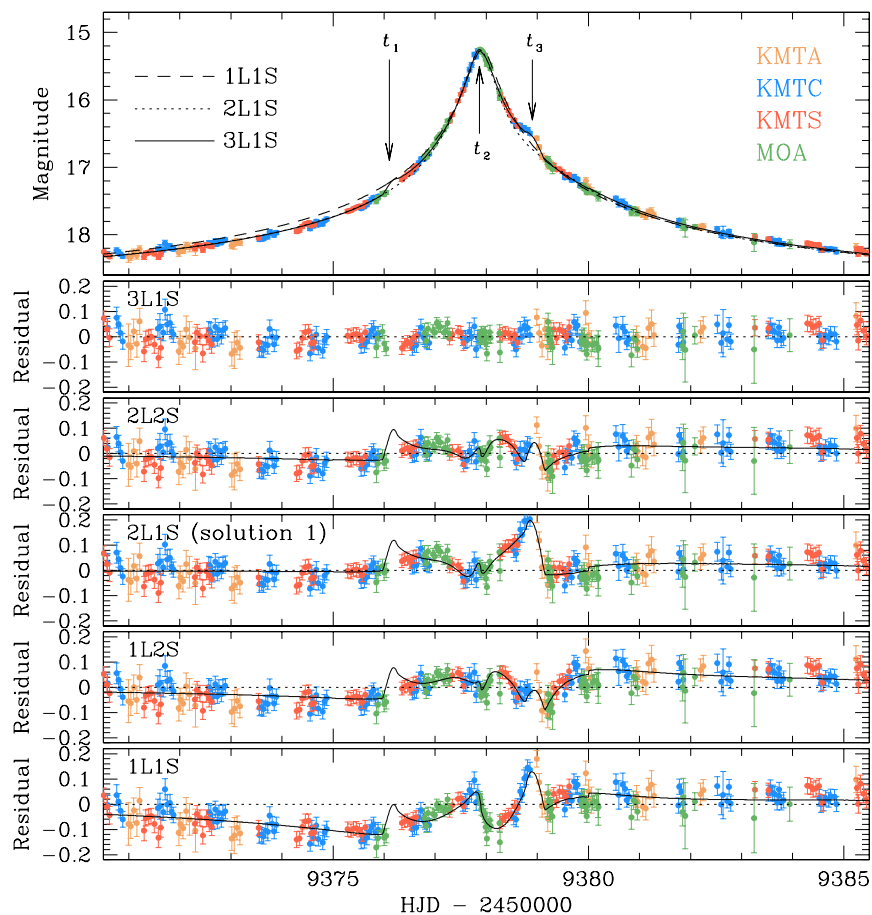}
\caption{
Zoom-in view around the peak of the light curve. The lower five panels show the residuals of 
the tested models under 3L1S, 2L2S, 2L1S, 1L2S, and 1L1S interpretations.  The curve drawn in each 
residual panel is the difference from the 3L1S model.  The curves of the 1L1S, 2L1S, and 3L1S 
models are drawn over the data points of the light curve in the top panel.  The epochs marked 
$t_1$, $t_2$, and $t_3$ indicate the times of the major anomalous features.  
}
\label{fig:two}
\end{figure*}

Because the event reached a high magnification at the peak, around which the light curve is
susceptible to anomalies induced by planetary companions \citep{Griest1998}, it was inspected 
after the peak was covered. A weak anomaly was seen to reside around the peak, but the 
deviations from a single-lens single-source (1L1S) model were not only weak but also smooth 
with few symptoms of caustic-involved features. As a result, no detailed analysis was 
carried out until we conducted a thorough reinvestigation of the light curve with improved 
photometry data acquired from optimized reduction of the images.

Observations by the KMTNet and MOA surveys were made mainly in the $I$ and MOA-$R$ bands, 
respectively, and some data were acquired in the $V$ band for the source color 
measurement. Reductions of the data were carried out using the photometry pipelines of the individual 
groups. Both of these pipelines, developed by \citet{Albrow2009} for the KMTNet survey and 
\citet{Bond2001} for the MOA survey, use the difference image technique \citep{Tomaney1996, 
Alard1998}, which is optimized for the photometry of stars lying in very dense star fields. Because 
the error bars from the pipelines tend to be underestimated, we renormalize the error bars 
so that they become consistent with the scatter of data and the $\chi^2$ per degree of freedom 
for the individual data sets becomes unity by applying the \citet{Yee2012} method.

\section{Interpreting the anomaly}\label{sec:three}

Figure~\ref{fig:one} shows the lensing light curve of KMT-2021-BLG-1077 constructed by combining 
the KMTNet and MOA data sets. Drawn over the data points is the 1L1S model obtained by fitting the 
data with the exclusion of those exhibiting deviations in the regions $9370.0\lesssim {\rm HJD}^\prime
\lesssim 9376.1$ and $9378.2 \lesssim {\rm HJD}^\prime \lesssim 9379.4$. The 1L1S lensing parameters 
are $(t_0, u_0, \te)\sim (9377.92, 0.012, 21.2)$.  Here the lensing parameters denote 
the time of the closest lens--source approach (expressed in HJD$^\prime$), 
the lens--source separation normalized to the angular Einstein radius $\thetae$ at that time (impact parameter),
 and the event 
timescale (expressed in days),
respectively.  Figure~\ref{fig:two} shows a zoomed-in view 
of the peak region and the residuals from the 1L1S model (bottom panel) to show the detailed 
features of the anomaly. This shows that the anomaly exhibits a complex deviation pattern, which 
is characterized by negative deviations in the region $\lesssim 9376.1$ ($t_1$) and two bumps 
in the residual at ${\rm HJD}^\prime\sim 9477.9$ ($t_2$) and $\sim 9478.9$ ($t_3$).  For an
explanation of the anomaly, we test models in conjunction with various interpretations of the 
lens and source configuration.

\subsection{Three-body models: 1L2S and 2L1S models}\label{sec:three-one}

We first check whether or not the deviations from the 1L1S model can be explained by three-body (lens plus
source) models, in which either the source or the lens is a binary. Hereafter, we denote lensing
events associated with a binary source \citep{Gaudi1998a, Dominik1998, Han1998} and a binary lens 
\citep{Mao1991} as 1L2S and 2L1S events, respectively.

Modeling the lensing light curve of a 1L2S event requires the inclusion of three extra lensing 
parameters in addition to those of the 1L1S modeling, $(t_{0,2}, u_{0,2}, q_F)$, in which the 
first two represent the approach time and impact parameter of the second source, $S_2$, 
to the lens, respectively, and the last parameter represents the flux ratio between the binary 
source stars. We denote the parameters related to the primary source, $S_1$, 
as $(t_{0,1}, u_{0,1})$. In order to consider finite-source effects, which arise when the lens 
passes over the surface of either of the source stars, we additionally include the normalized 
source radii $\rho_1$ and $\rho_2$, which are defined as the ratios of the angular source radii, 
$\theta_{*,1}$ and $\theta_{*,2}$, to $\thetae$, that is, $\rho_1 =\theta_{*,1}/\thetae$ and $\rho_2 
=\theta_{*,2}/\thetae$.  The modeling was conducted by checking various trajectories of the second 
source with the consideration of the anomalous features in the 1L1S residual. For a given set of 
the initial lensing parameters related to $S_1$, the best-fit 1L2S solution was searched for by 
minimizing $\chi^2$ using a downhill approach based on the Markov Chain Monte Carlo (MCMC) 
algorithm.  The 1L2S modeling yields a model with lensing parameters of $(t_{0,1}, u_{0,1}, t_{0,2}, 
u_{0,2}, \te, q_F)\sim (9377.875, 8.69\times 10^{-3}, 9378.807, 8.98\times 10^{-3}, 25.6, 0.104)$, 
and the normalized radii of both source stars are not constrained.

The residual from the best-fit 1L2S model is presented in Figure~\ref{fig:two}. It shows that 
the model substantially reduces the residual from the 1L1S model by $\Delta\chi^2=531.3$, 
but the model still leaves noticeable residuals throughout the peak region.  The residuals are 
characterized by negative deviations at a 0.05~mag level before $t_1$, positive deviations of a 
similar level after $t_3$, and a wiggly pattern with alternating positive and negative deviations 
in the region around $t_2$. This indicates that the 1L2S model is not a correct interpretation of 
the anomaly.

\begin{figure}[t]
\includegraphics[width=\columnwidth]{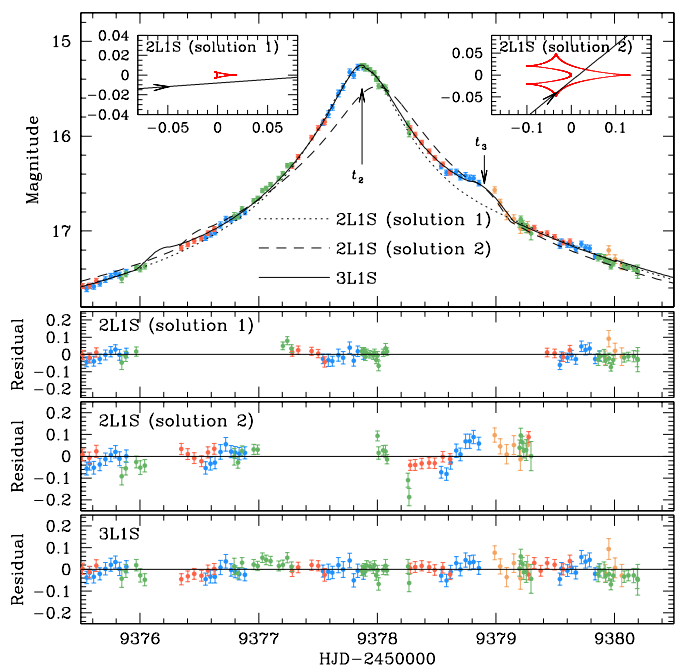}
\caption{
Models and residuals of the ``solution~1'' and ``solution~2'' 2L1S models.  The individual 
solutions are obtained from the two sets of 2L1S models, in which the light curve was fitted 
by excluding the data lying in the regions of
$9376.0 < {\rm HID}^\prime < 9377.2$ and
$9378.2 < {\rm HID}^\prime < 9379.4$ for solution~1, and
by excluding those lying in the regions of
$9377.0 < {\rm HID}^\prime < 9378.0$ and
$9379.3 < {\rm HID}^\prime < 9381.0$ for solution~2.
The two insets in the top panel show the lens system configurations of the individual solutions.  
Also presented are the model curve and residual of the 3L1S solution.
}
\label{fig:three}
\end{figure}

Similar to the 1L2S case, a 2L1S modeling also requires three additional parameters to 
describe the lens binarity. These are $(s, q, \alpha)$, which denote the projected separation 
(scaled to $\thetae$) and mass ratio between the binary lens components, $M_1$ and $M_2$, and 
the source trajectory angle, which is defined as the angle between the direction of the source--lens 
relative motion and the binary axis, respectively.  For this event, it is also necessary to 
include the normalized source size $\rho=\theta_*/\theta_{\rm E}$ in order to take into 
consideration finite-source effects.  In the 2L1S modeling, we find a lensing solution in two 
steps, in which the binary parameters $s$ and $q$ were searched for via a grid approach and the 
other parameters were found via a downhill approach in the first step, and then local solutions
found from the first step were polished by releasing all parameters as free parameters in the 
second step.  It is found that the 2L1S interpretation does not yield a model describing all 
the features of the anomaly either.

We then checked whether or not a 2L1S model could partially explain the anomaly.  We test this 
possibility because the central anomaly can be affected by an additional lens companion, 
if it exists, and in this case, a 2L1S model can often describe a part of the anomaly.  
For this test, we conducted two sets of 2L1S modeling in which the light curve was fitted 
by excluding the data lying in the regions of
$9376.0 < {\rm HID}^\prime < 9377.2$ and
$9378.2 < {\rm HID}^\prime < 9379.4$ in the first modeling, and
by excluding those lying in the regions of
$9377.0 < {\rm HID}^\prime < 9378.0$ and
$9379.3 < {\rm HID}^\prime < 9381.0$ in the second modeling.
We designate the solutions found from the modeling of the individual data sets as ``solution~1'' 
and ``solution~2'', respectively.  In Figure~\ref{fig:three}, we present the model curves and 
residuals of solutions~1 and 2.  The model and residuals of solution~1 showing all data 
are separately presented in Figure~\ref{fig:two}.  Figure~\ref{fig:three} shows that solution~1 nicely explains the anomaly feature around the peak centered at $t_1$, while solution~2 describes the bump around $t_3$.  The lensing parameters of the models are $(s, q, 
\alpha, \rho) \sim (1.31 (1/1.31), 1.6\times 10^{-3}, 3.06, 5.4\times 10^{-3})$ for solution~1, and $\sim (0.98, 1.2\times 10^{-3}, 2.46, 4.2\times 10^{-3})$ for solution~2.  
Here, the source trajectory angle is  expressed in radians.  For solution~1, we note that 
there exist two locals with binary separations of $s\sim 1.31$ and $s\sim 1/1.31$, for which 
the similarity in the lensing light curves between the two local solutions is caused by the 
well-known close--wide degeneracy arising due to the similarity between the central lensing 
caustics formed by binary lenses with $s$ and $1/s$ \citep{Griest1998, Dominik1999, An2005}.  
Here, the caustics represent the source positions at which the lensing magnification of a point 
source becomes infinite.  The two insets presented in the top panel of Figure~\ref{fig:three} 
show the lens system configurations of the individual solutions.

\subsection{Four-body models: 2L2S and 3L1S models}\label{sec:three-two}

Recognizing the inadequacy of the three-body models, we then examine two four-body models. In one of these, both the source and lens are binaries (2L2S model), and in the other, the lens 
is a triple system (3L1S model). In the 2L2S modeling, we add three binary-source parameters $(t_{0,2}, u_{0,2}, q_F)$ to those of the 2L1S model. In the 3L1S modeling, we add three 
tertiary-lens parameters $(s_3, q_3, \psi)$, where $s_3$ and $q_3$ represent the separation 
and mass ratio between the third lens component, $M_3$, and the primary, and $\psi$  
denotes the orientation angle of $M_3$ as measured from the $M_1$--$M_2$ axis with a center 
at the position of $M_1$. We add the subscript ``2'' to the lensing parameters describing 
$M_2$, that is, $(s_2, q_2)$, to distinguish them from those describing $M_3$.

The 2L2S modeling was carried out by checking various trajectories of the second source under 
the basic lens-system configurations of the 2L1S models,  that is, solutions 1 and 2 of 
the 2L1S models.  In Figure~\ref{fig:two}, we present the residual of the best-fit 2L2S model 
found based on solution~1 of the 2L1S model. The binary-source parameters of the solution 
are $(t_{0,2}, u_{0,2}, q_F)\sim (9378.811, 5.74\times 10^{-3}, 0.042)$, and the other 
parameters are very similar to those of the 2L1S model. From inspection of the residual, we find that the model substantially reduces the 2L1S residuals in the region between 
$t_1$ and $t_3$ and improves the fit by $\Delta\chi^2=327.1$ with respect to the 2L1S model. 
However, the model still leaves subtle wiggles in the residual throughout the peak region.  
It is found that the 2L2S model obtained based on solution~2 of the 2L1S model results 
in a poorer fit than the model found based on solution~1.

\begin{figure}[t]
\includegraphics[width=\columnwidth]{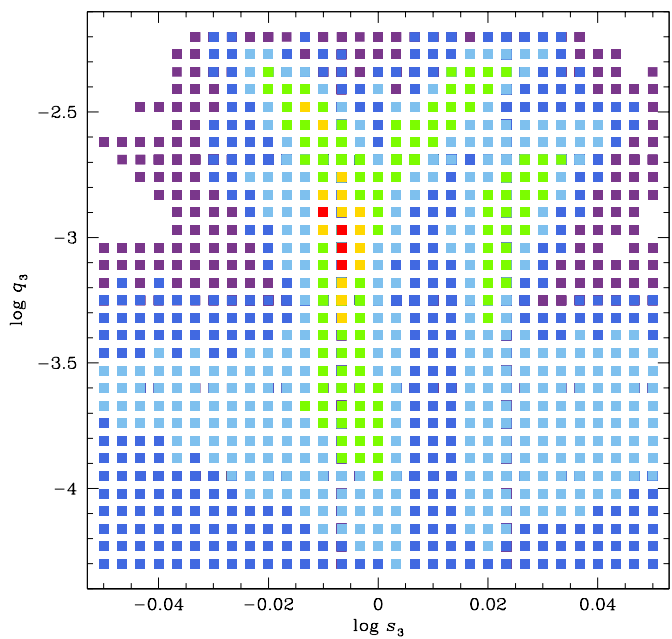}
\caption{
Distribution of $\Delta\chi^2$ on the $\log s_3$--$\log q_3$ parameter plane. The color coding 
is arranged to denote regions with 
$\leq 1n\sigma$ (red), 
$\leq 2n\sigma$ (yellow), 
$\leq 3n\sigma$ (green), 
$\leq 4n\sigma$ (cyan), 
$\leq 5n\sigma$ (blue), and 
$\leq 6n\sigma$ (purple), where $n=3$.
}
\label{fig:four}
\end{figure}

We further checked the 3L1S interpretation of the anomaly.  Similar to the 2L1S modeling, the 
3L1S modeling was carried out in two steps.  In the first round, we inspected the space of 
the tertiary lens parameters, that is, $(s_3, q_3, \psi)$, via a grid approach by fixing the 
other lensing parameters as the values found from the 2L1S modeling. In the second round, we 
refined the locals found in the $s_3$--$q_3$--$\psi$ planes by releasing all parameters as 
free parameters. This approach is based on the fact that the anomalies induced by two 
companions to the lens can, in many cases, be approximated by the superposition of the anomalies 
induced by the $M_1$--$M_2$ and $M_1$--$M_3$ binary pairs \citep{Bozza1999, Han2001}.

Figure~\ref{fig:four} shows the $\Delta\chi^2$ distribution on the $\log s_3$--$\log q_3$ 
parameter plane obtained from the grid searches for these parameters in the first round of 
the 3L1S modeling conducted based on solution 1 of the 2L1S model.  It shows a single 
distinct local at $(\log s_3, \log q_3)\sim (-0.01, -3,0)$. The full lensing parameters of 
the 3L1S models after refining the local through the second round of the modeling are listed 
in Table~\ref{table:one}.  We note that $\alpha$, that is, the angle between the source trajectory 
and the $M_1$--$M_2$ axis, is similar for the 3L1S model and for the 2L1S solution 1, while 
$(\alpha +\psi -2\pi)$, that is, the angle between the source trajectory and the $M_1$--$M_3$ axis 
(additionally inserted in Table~\ref{table:one}), for the 3L1S model is similar to the $\alpha$ 
for the 2L1S solution 2.  We find two sets of 3L1S solutions, which are obtained from the modeling 
based on the close and wide 2L1S models of solution~1.  We refer to the individual solutions 
as close and wide 3L1S solutions, between which the wide solution is favored over the close 
solution by $\Delta\chi^2 = 43.4$.  It is found that the 3L1S model found on the basis of 
solution~2 of the 2L1S model is consistent with the 3L1S model obtained based on solution~1, 
except that the order of $M_2$ and $M_3$ is reversed.

\begin{table}[t]
\small
\caption{Lensing parameters of the 3L1S models\label{table:one}}
\begin{tabular*}{\columnwidth}{@{\extracolsep{\fill}}lcccc}
\hline\hline
\multicolumn{1}{c}{Parameter}    &
\multicolumn{1}{c}{Close}        &
\multicolumn{1}{c}{Wide }        \\
\hline
$\chi^2$                &   $1547.1              $   &  $1503.7               $   \\
$t_0$ (HJD$^\prime$)    &   $9377.971 \pm 0.003  $   &  $9377.9596 \pm 0.003  $   \\
$u_0$ ($10^{-3}$)       &   $-9.57 \pm 0.30      $   &  $-10.63 \pm 0.37      $   \\
$\te$ (days)            &   $28.64 \pm 0.66      $   &  $24.85 \pm 0.73       $   \\
$s_2$                   &   $0.767 \pm 0.005     $   &  $1.311 \pm 0.009      $    \\
$q_2$ ($10^{-3}$)       &   $1.62 \pm 0.10       $   &  $1.56 \pm 0.11        $    \\
$\alpha$ (rad)          &   $3.062 \pm 0.003     $   &  $3.068 \pm 0.003      $    \\
$s_3$                   &   $0.979 \pm 0.001     $   &  $0.973 \pm 0.002      $    \\
$q_3$ ($10^{-3}$)       &   $1.30 \pm 0.09       $   &  $1.75 \pm 0.16        $    \\
$\psi$ (rad)            &   $5.702 \pm 0.034     $   &  $5.733 \pm 0.036      $    \\
$\alpha+\psi-2\pi$      &    2.481                   & 2.518                       \\
$\rho$ ($10^{-3}$)      &   $4.82 \pm 0.31       $   &  $5.40 \pm 0.33        $    \\
\hline
\end{tabular*}
\tablefoot{ ${\rm HJD}^\prime = {\rm HJD}- 2450000$.  }
\end{table}

It is found that the triple-lens interpretation yields a model that can explain all the major anomalous 
features in the lensing light curve.  This can be seen in the model curve and the residual from 
the model presented in  Figures~\ref{fig:two} and \ref{fig:three}.  We find that the wide 3L1S 
model yields a fit that is better than the 1L1S, 1L2S, 2L1S, and 2L2S models by $\Delta\chi^2=937.0$, 
405.7, 486.4, and 159.3, respectively. In the residual panels of the other models in Figure~\ref{fig:two}, 
we present the curves of the differences from the 3L1S model. The estimated mass ratios of the lens 
companions to the primary are $q_2=M_2/M_1\sim 1.56 \times 10^{-3}$ and $q_3=M_3/M_1\sim 1.75\times 
10^{-3}$.  These mass ratios correspond to $\sim 1.6$ and $\sim 1.8$ times the Jupiter/Sun mass 
ratio, respectively, and therefore the lens is a multiplanetary system containing two giant planets.

\begin{figure}[t]
\includegraphics[width=\columnwidth]{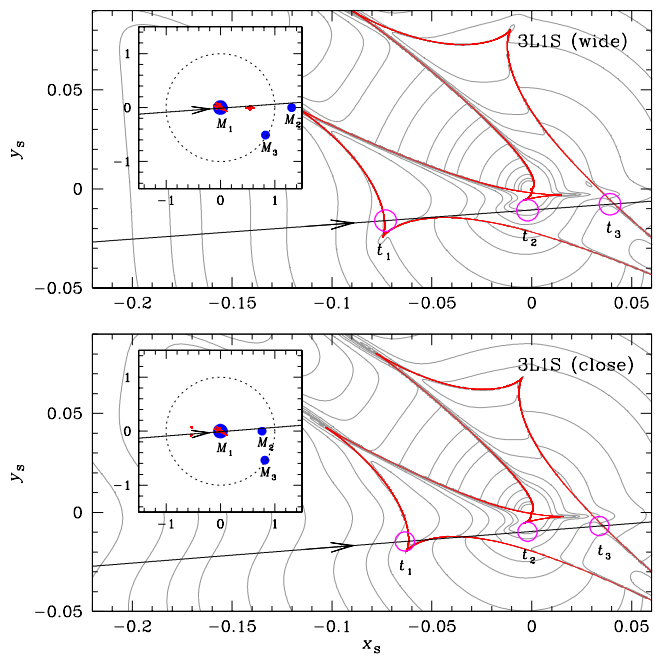}
\caption{
Configuration of the lens system according to the wide (upper panel) and close (lower panel) 
3L1S models. In each panel, the cuspy figure is the caustic and the line with an arrow 
indicates the source trajectory. The three empty magenta circles on the source trajectory 
represent the source positions corresponding to the three epochs of $t_1$, $t_2$, and $t_3$ 
marked in Fig.~\ref{fig:one}. The size of the circle is proportionate to the source size. Lengths 
are normalized to the Einstein radius corresponding to the total lens mass. The inset in each 
panel shows a zoomed-out view, in which the locations of the individual lens components ($M_1$, 
$M_2$, and $M_3$) and the Einstein ring (dotted circle) are marked.  The curves encompassing 
the caustic represent the equi-magnification contours.
}
\label{fig:five}
\end{figure}

Figure~\ref{fig:five} displays the lens system configuration, which shows the source trajectory 
(line with an arrow) relative to the positions of the lens components (blue dots marked by $M_1$, 
$M_2$, and $M_3$) and caustics (red cuspy figure).  Although the close solution is disfavored with 
a significant $\chi^2$ difference from the wide solution, we present its lens system configuration 
for comparison with that of the wide solution. For each solution, the main panel shows a 
zoomed-in view of the central magnification region, and the inset shows the wider view encompassing 
all the lens components.  The central caustic appears to be the combination of a tiny wedge-shaped 
central caustic induced by $M_2$ and a bigger resonant caustic induced by $M_3$.  The similarities 
of the individual caustics to those of the 2L1S caustics of solutions 1 and 2, shown in the 
insets of the top panel of Figure~\ref{fig:three}, indicate that the anomaly can be approximated 
by the superposition of the anomalies induced by the two planetary companions.  According to the 
3L1S interpretation, the negative deviation from the 1L1S model before $t_1$ was produced by the 
source passage through the negative-deviation region extending from the back end of the $M_2$-induced 
caustic.  The source entered the resonant caustic induced by $M_3$ at around $t_1$, and passed along 
one of the caustic folds before it crossed the tip of the $M_1$-induced caustic at around $t_2$, 
which corresponds to the 1L1S residual bump at the corresponding time. The source further proceeded 
and exited the resonant caustic at $t_3$, and this produced the second bump in the 1L1S residual at 
the corresponding time.  We mark the source positions corresponding to the epochs of the major 
anomalous features at $t_1$, $t_2$, and $t_3$ with empty magenta circles, whose size is proportionate to 
the source size.

\begin{figure}[t]
\includegraphics[width=\columnwidth]{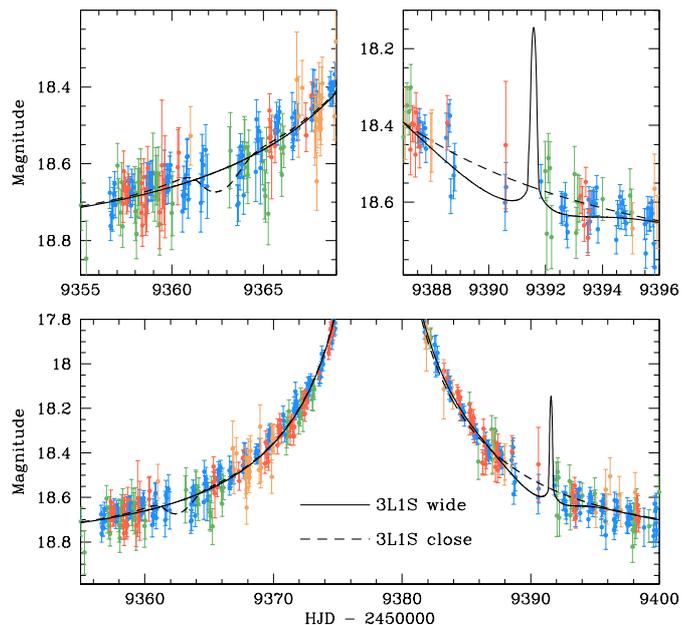}
\caption{
Enlarged view of the light curve around the regions of the planetary-caustic induced anomalies 
predicted by the close (upper left panel) and wide (upper right panel) models. The bottom panel 
shows a zoomed-out view of the light curve at low magnification.  The solid and dashed curves 
drawn over the data points are the wide and close 3L1S models, respectively.
}
\label{fig:six}
\end{figure}

According to the 3L1S solution, the passage of the source was almost parallel with the $M_1$--$M_2$ 
axis, with a source trajectory angle of $\alpha\sim 4.6^\circ$. This indicates the possibility that 
the source additionally approached the peripheral (planetary) caustic induced by $M_2$, and we find 
that the source passed the region around the planetary caustic according to either the 
close or the wide model, roughly 15 days before or after the peak, respectively.  See the insets  
of the panels presented in Figure~\ref{fig:five}.  If an additional anomaly were produced by this 
approach and captured by the data, it would further constrain the 3L1S interpretation. We therefore
checked the data around the time of the anomaly expected by the individual models. Figure~\ref{fig:six} 
shows the light curve around the regions of the anomalies predicted by the close (upper left panel) 
and wide (upper right panel) solutions.  The light curve would exhibit a short-term dip and a bump 
at around ${\rm HJD}^\prime \sim 9362.5$ and $\sim 9391.5$ according to the close and wide solutions, 
respectively.  However, we find that it was difficult to confirm the additional anomaly in the light 
curve due to the large photometric uncertainties of the data around the regions, which   lie close to 
the baseline.  We note that the difficulty in identifying an extra anomaly may be caused by the orbital 
motion of the first planet, that is, $M_2$. According to the Bayesian analysis of the physical lens 
parameters, which we discuss in Sect.~\ref{sec:five}, the host mass is about $0.14~M_\odot$, and the 
projected separation of the planet is about 0.8~AU and 1.4~AU according to the close and wide solutions, 
respectively.  Hence in 15~days, it is expected for the orbital angle to change 
by $\sim 6^\circ$ and $2^\circ\hskip-2pt .5$ according to the individual solutions. 
Consequently, even if the static model predicted an extra anomaly, it might not appear 
in the light curve because the planet had moved due to its orbital motion.

\section{Source star and angular Einstein radius}\label{sec:four}

For the lensing event KMT-2021-BLG-1077, it is possible to measure the extra observable of 
the angular Einstein radius because the anomaly in the lensing light curve was affected by 
finite-source effects.  With the normalized source radius measured from the analysis of the 
anomaly, the Einstein radius is estimated as
\begin{equation}
\thetae={\theta_* \over \rho},
\label{eq1}
\end{equation}
where $\theta_*$ is the angular radius of the source star.  With the measured $\thetae$, 
together with the basic observable of the event timescale, the physical parameters of the 
lens mass, $M$, distance to the lens, $\dl$, and lens--source relative proper motion $\mu$
can be constrained using the relations
\begin{equation}
\mu = {\thetae\over \te}; \ \ \ 
\thetae = (\kappa M \pi_{\rm rel})^{1/2};\ \ \ 
\pi_{\rm rel} = {\rm AU}\left( {1\over D_{\rm L}} - {1 \over D_{\rm S}}\right),
\label{eq2}
\end{equation}
where $\kappa =4G/(c^2{\rm AU})$ and $D_{\rm S}$ indicates the distance to the source. 
If an additional lensing observable of the microlens-parallax $\pivec_{\rm E}=(\pi_{\rm rel}/
\thetae)(\muvec/\mu)$ can be measured, the lens mass and distance are uniquely determined by
\begin{equation}
M = {\thetae \over \kappa\pie };\qquad
\dl = { {\rm AU} \over \pie\thetae + \pi_{\rm S} },
\label{eq3}
\end{equation}
where $\pi_{\rm S}={\rm AU}/D_{\rm S}$ is the parallax of the source \citep{Gould1992, Gould2000}. 
The microlens parallax can sometimes be measured from the subtle deviations in the lensing light 
curve caused by the positional change of an observer induced by the orbital motion of Earth around 
the Sun \citep{Gould1992}.  We find that it is difficult to securely constrain $\pie$ because 
the photometric precision of the data in the wings of the light curve is not 
sufficiently
high to 
detect subtle deviations induced by the microlens-parallax effect.

\begin{figure}[t]
\includegraphics[width=\columnwidth]{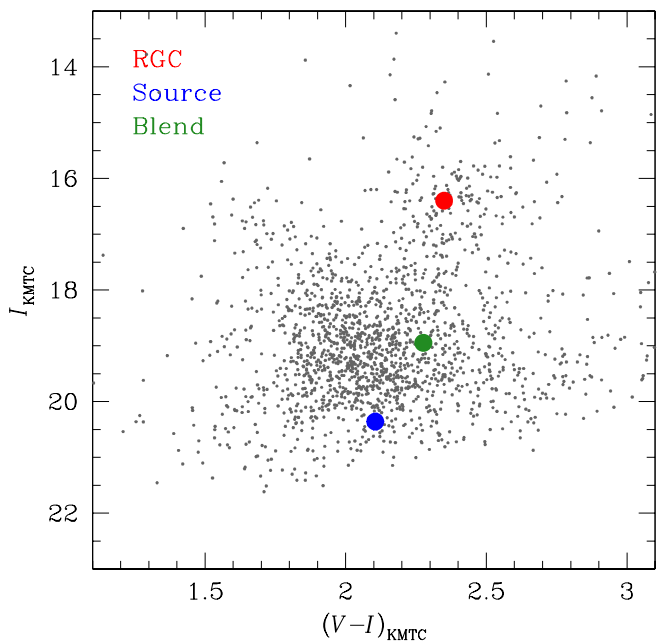}
\caption{
Locations of the source (blue dot) and centroid of the red giant clump (red dot) on the 
instrumental color--magnitude diagram of stars lying around the source, constructed from the 
pyDIA photometry of the KMTC data set. Also marked is the location of the blend (green dot).
}
\label{fig:seven}
\end{figure}

For the measurement of $\thetae$, we first estimate the angular source radius, which is deduced 
from the color and magnitude of the source. Figure~\ref{fig:seven} shows the location of the source 
(blue solid dot) in the instrumental color--magnitude diagram (CMD) of stars lying in the vicinity 
of the source constructed from the pyDIA photometry \citep{Albrow2017} of the KMTC data. Also 
marked is the location of the blend (green dot). As discussed in Sect.~\ref{sec:five}, 
the blend is not the lens but a star (or stars) not involved with the lensing magnification. The 
$I$- and $V$-band source magnitudes were determined from the regression of the pyDIA data with 
the variation of the lensing magnification. Following the \citet{Yoo2004} method, we calibrate 
the source color and magnitude using the centroid of the red giant clump (RGC) marked 
by a red dot in Figure~\ref{fig:seven} as a reference.  From the offsets in color and magnitude 
between the source and RGC centroid, $\Delta (V-I, I)$, together with the known reddening and 
extinction-corrected (de-reddened) values of the RGC centroid, $(V-I, I)_{{\rm RGC},0}=(1.060, 14.607)$ 
\citep{Bensby2013, Nataf2013}, we estimate the de-reddened source color and magnitude as
\begin{equation}
\eqalign{
(V-I, I)_0 =  & (V-I, I)_{{\rm RGC},0} + \Delta(V-I, I) \cr
           =  &  (0.815\pm 0.007, 18.564\pm 0.001),     \cr
}
\label{eq4}
\end{equation}
indicating that the source is a late G-type main sequence star.

Once $(V-I, I)_0$ are measured, we convert $V-I$ color into $V-K$ color using the color--color 
relation of \citet{Bessell1988}, and then interpolate $\theta_*$ from the \citet{Kervella2004} 
relation between $V-K$ and $\theta_*$.  The angular radius of the source estimated from this 
procedure is 
\begin{equation}
\theta_* = 0.69 \pm 0.048~\mu{\rm as}.
\label{eq5}
\end{equation}
The Einstein radius is then estimated from the relation in Equation~(\ref{eq1}) as
\begin{equation}
\thetae = 0.13 \pm 0.01~{\rm mas},
\label{eq6}
\end{equation}
and the relative lens-source proper motion is estimated using the measured event time scale as
\begin{equation}
\mu = {\thetae\over \te} = 1.81 \pm 0.13~{\rm mas}~{\rm yr}^{-1}.
\label{eq7}
\end{equation}
We note that the values of $\thetae$ and $\mu$ are estimated using the lensing parameters 
of the wide 3L1S solution because it is favored over the close solution with a significant 
level, that is, $\Delta\chi^2=43.4$.  The estimated Einstein radius is substantially smaller 
than $\sim 0.5$~mas of a typical lensing event produced by a low-mass star with a mass of 
$\sim 0.3~M_\odot$ lying roughly halfway between the source and observer.  This suggests that 
the lens would either have a very low mass or lie close to the source.

\begin{figure}[t]
\includegraphics[width=\columnwidth]{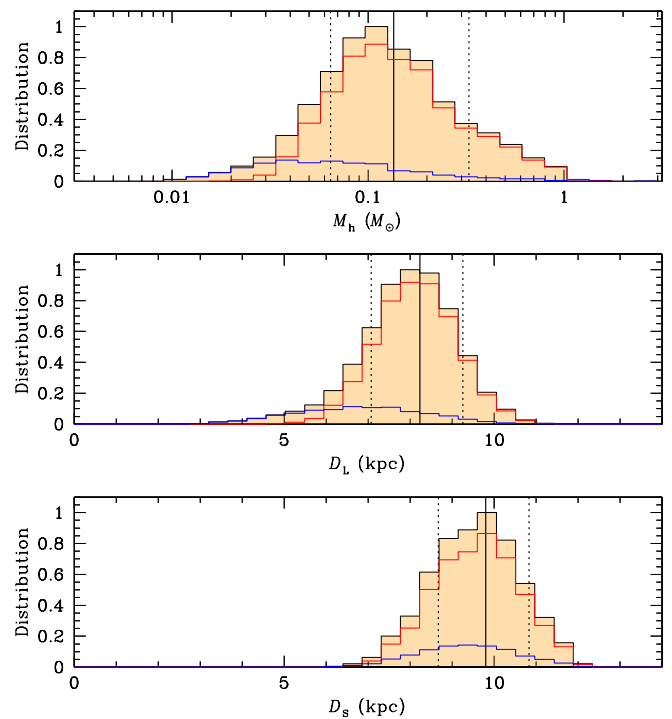}
\caption{
Bayesian posteriors of the host mass ($M_{\rm h}$), distances to the planetary system ($\dl$), 
and source ($D_{\rm S}$). In each panel, the blue and red curves represent the contributions 
by the disk and bulge lenses, respectively, and the black curve is the sum of the two lens 
populations. The solid vertical line represents the median of the distribution, and the two 
dotted lines represent the 1$\sigma$ range.
}
\label{fig:eight}
\end{figure}

\section{Physical parameters of the planetary system}\label{sec:five}

In this section, we estimate the physical parameters of the planetary system using the
constraints provided by the measured observables of $\te$ and $\thetae$. Not being able to
uniquely constrain $M$ and $\dl$ using the relations in Equation~(\ref{eq3}) due to the 
absence of a $\pie$ constraint, we statistically estimate the parameters by conducting a 
Bayesian analysis using a Galactic model.  The analysis is done based on the observables 
of the wide 3L1S solution.

The Galactic model defines the physical and dynamical distributions of stars and remnants
in the Galaxy and their mass function (MF). We use the Galactic model of \citet{Jung2021}, in 
which \citet{Robin2003} and \citet{Han2003} models are used for the physical distributions 
of the disk and bulge objects, respectively; \citet{Jung2021} and \citet{Han1995} models are 
adopted for the dynamical distributions of the disk and bulge objects, respectively; and 
the \citet{Jung2018} MF is employed for the MF of both disk and bulge objects.  The MF 
model was constructed by adopting the initial MF and the present-day MF of \citet{Chabrier2003} 
for the bulge and disk lens populations, respectively.

In the first step of the Bayesian analysis, we generate a large number ($10^7$) of artificial 
lensing events by conducting a Monte Carlo simulation using the Galactic model. For the 
individual simulated events, we then compute the values of the observables $\te$ and 
$\thetae$ using the relations in Equation~(\ref{eq2}).  In the second step, we construct 
the probability distributions of events by imposing Gaussian weight based on the measured 
values of the observables, that is, $\te=(24.85\pm 0.73)$~days and $\thetae=(0.13\pm 0.01)$~mas.

Figure~\ref{fig:eight} shows the Bayesian posteriors of the host mass, $M_{\rm h}$, and the 
distances to the planetary system and the source star.  In each distribution, we mark the 
median and uncertainty (1$\sigma$) range of the distribution by a solid vertical line and 
dotted lines, respectively. The uncertainty range is estimated as the 16\% and 84\% of the 
distribution. The curves marked in blue and red indicate the disk and bulge lens contributions, 
respectively, and the black curve represents the sum of the contributions from the two lens 
populations.  It is found that the relative probabilities of the disk and bulge lenses are 
17\% and 83\%, respectively.  We present the distribution of $D_{\rm S}$ to show the relative 
positions of the lens and source.  The distributions of $\dl$ and $D_{\rm S}$ indicate that 
the event is likely to be produced by a star lying in the bulge, and the closeness between the 
lens and source explains the reason for the small Einstein radius. See the relation between 
$\thetae$ and $(D_{\rm S},D_{\rm L})$ in Equation~(\ref{eq2}).

It is found that KMT-2021-BLG-1077L is a planetary system, in which a mid-to-late M dwarf star 
hosts two gas giant planets with masses of slightly less than that of Saturn in our Solar System.  
The estimated mass of the planet host is 
\begin{equation}
M_{\rm h} = 0.14^{+0.19}_{-0.07}~M_\odot, 
\label{eq8}
\end{equation}
and the masses of the two planets are 
 \begin{equation}
M_{\rm p_1} = 0.22^{+0.31}_{-0.12}~M_{\rm J}, \qquad
M_{\rm p_2} = 0.25^{+0.35}_{-0.13}~M_{\rm J}.
\label{eq9}
\end{equation}
The planetary system is likely to be in the bulge with a distance from Earth of 
\begin{equation}
\dl = 8.24^{+1.02}_{-1.16}~{\rm kpc}.
\label{eq10}
\end{equation}
The projected separations of the individual planets from the host are
\begin{equation}
a_{\perp, {\rm p}_1} = 1.26^{+1.41}_{-1.08}~{\rm AU},\qquad
a_{\perp, {\rm p}_2} = 0.93^{+1.05}_{-0.80}~{\rm AU}.
\label{eq11}
\end{equation}
Considering that the snow line distance is $d_{\rm sl} = 2.7(M_{\rm h}/M_\odot)\sim 0.38$~AU
and the separations are projected ones, both planets lie well beyond the snow line of the host. 
If we assume that $a = a_\perp$, then at face value, the system is not stable. However,
$a_\perp$ is simply the minimum possible value for the semi-major axis, and a stable system
can easily be achieved simply by moving one or the other planet in or out of the plane of the
sky. \citet{Madsen2019} explored the issue of stability for the OGLE-2012-BLG-0026L system.
The ideas discussed in that paper are broadly applicable to systems such as
KMT-2021-BLG-1077L, which appear at first glance to be unstable based on their projected
separations.

The estimated host mass and distance indicate that the contribution of the lens flux to the 
blend, marked on the CMD in Figure~\ref{fig:seven} (green solid dot), is negligible. This is 
additionally confirmed by the astrometric measurement of the offset between the source, 
measured on the difference image obtained during the lensing magnification, and the baseline 
object in the reference image. The offset, $\sim 0.352$~arcsec, is far larger than the 
measurement error, which is of the order of 10~mas.

The lens of the event KMT-2021-BLG-1077 is the fifth confirmed multiplanetary system 
detected by microlensing following  OGLE-2006-BLG-109L, OGLE-2012-BLG-0026L, OGLE-2018-BLG-1011L, 
and OGLE-2019-BLG-0468L.  It is the third multiplanetary system detected after the full operation 
of the high-cadence surveys.  If the lenses of the events KMT-2019-BLG-1953 and KMT-2021-BLG-0240 ---
which exhibit relatively less-secure multi-planet signatures--- are multiplanetary systems, the 
total number of  multiplanetary systems detected by the six-year operation of the high-cadence 
surveys is five.  Considering that about 20 planets are annually detected by the surveys, this 
roughly matches the prediction of \citet{Zhu2014} that $\sim 5.5\%$ of planetary events that 
are detectable by the high-cadence surveys will exhibit multi-planet signals, although the 
number is too small to draw strong conclusions.

\section{Summary and conclusion}\label{sec:six}

We present an analysis of the high-magnification microlensing event KMT-2021-BLG-1077, 
for which the peak region of the lensing light curve exhibited a subtle and complex anomaly 
pattern.  The anomaly could not be explained by the usual three-body models, in which either the 
lens or the source is a binary.  However, the anomaly can be explained by a lensing model in which 
the lens is composed of three masses.

With the constraints of the event timescale and angular Einstein radius, it is found that 
the lens is a multiplanetary system residing in the Galactic bulge.  The primary of the lens 
system is a mid-to-late M dwarf and hosts two gas giant planets lying  beyond the snow line 
of the host.  The lens of the event is the fifth confirmed multiplanetary system detected using 
microlensing following OGLE-2006-BLG-109L, OGLE-2012-BLG-0026L, OGLE-2018-BLG-1011L, and 
OGLE-2019-BLG-0468L.

\begin{acknowledgements}
Work by C.H. was supported by the grants  of National Research Foundation of Korea 
(2020R1A4A2002885 and 2019R1A2C2085965).
This research has made use of the KMTNet system operated by the Korea Astronomy and Space 
Science Institute (KASI) and the data were obtained at three host sites of CTIO in Chile, 
SAAO in South Africa, and SSO in Australia.
he MOA project is supported by JSPS KAKENHI grant Nos. JSPS24253004,
JSPS26247023, JSPS23340064, JSPS15H00781, JP16H06287, JP17H02871, and JP19KK0082. 
J.C.Y.  acknowledges support from NSF Grant No. AST-2108414.
C.R. was supported by the Research fellowship of the Alexander von Humboldt Foundation.
\end{acknowledgements}


\begin{thebibliography}{}
\bibitem[Alard \& Lupton(1998)]{Alard1998} Alard, C., \& Lupton, R. H.\ 1998, \apj, 503, 325
\bibitem[Albrow(2017)]{Albrow2017} Albrow, M.\ 2017, MichaelDAlbrow/pyDIA: Initial Release on Github,Versionv1.0.0, Zenodo, doi:10.5281/zenodo.268049
\bibitem[Albrow et al.(2009)]{Albrow2009} Albrow, M., Horne, K., Bramich, D.~M., et al.\ 2009, \mnras, 397, 2099
\bibitem[An(2005)]{An2005} An, J. H. 2005, \mnras, 356, 1409
\bibitem[Beaulieu et al.(2016)]{Beaulieu2016}Beaulieu, J. -P., Bennett, D. P., Batista, V., et al. 2016, \apj, 824, 83
\bibitem[Beaulieu et al.(2006)]{Beaulieu2006} Beaulieu, J. -P., Bennett, D. P., Fouqu\'e, P., et al. 2006, Nature, 439, 437
\bibitem[Bennett et al.(2010)]{Bennett2010} Bennett, D. P., Rhie, S. H., Nikolaev, S., et al. 2010, \apj, 713, 837
\bibitem[Bensby et al.(2013)]{Bensby2013} Bensby, T., Yee, J.~C., Feltzing, S., et al.\ 2013, \aap, 549, A147
\bibitem[Bessell \& Brett(1988)]{Bessell1988} Bessell, M.~S., \& Brett, J.~M. 1988, \pasp, 100, 1134
\bibitem[Blackman et al.(2021)]{Blackman2021} Blackman, J. W., Beaulieu, J. P., Bennett, D. P., et al. 2021, Nature, 598, 272
\bibitem[Bond et al.(2001)]{Bond2001} Bond, I. A., Abe, F., Dodd, R. J., et al. 2001, \mnras, 327, 868
\bibitem[Bozza(1999)]{Bozza1999} Bozza, V. 1999, \aap, 348, 311
\bibitem[Chabrier(2003)]{Chabrier2003} Chabrier, G. 2003, \pasp, 115, 763
\bibitem[Dominik(1998)]{Dominik1998} Dominik, M.\ 1998, \aap, 333, 893
\bibitem[Dominik(1999)]{Dominik1999} Dominik, M.\ 1999, \aap, 349, 108
\bibitem[Gaudi(2012)]{Gaudi2012} Gaudi, B. S. 2012, \araa, 50, 411
\bibitem[Gaudi(1998)]{Gaudi1998a} Gaudi, B. S. 1998, ApJ, 506, 533
\bibitem[Gaudi et al.(2008)]{Gaudi2008} Gaudi, B. S., Bennett, D. P., Udalski, A., et al. 2008, Science, 319, 927
\bibitem[Gaudi et al.(1998)]{Gaudi1998b} Gaudi, B. S., Naber, R. M., \& Sackett, P. D. 1998, \apj, 502, L33
\bibitem[Gould(1992)]{Gould1992} Gould, A.\ 1992, \apj, 392, 442
\bibitem[Gould(2000)]{Gould2000} Gould, A.\ 2000, \apj, 542, 785
\bibitem[Gould et al.(2006)]{Gould2006} Gould, A., Udalski, A., An, D., et al. 2006, \apj, 644, L37
\bibitem[Griest \& Safizadeh(1998)]{Griest1998} Griest, K., \& Safizadeh, N.\ 1998, \apj, 500, 37
\bibitem[Han(2005)]{Han2005} Han, C. 2005, \apj, 629, 1102
\bibitem[Han(2019)]{Han2019} Han, C., Bennett, D. P., Udalski, A., et al. 2019, \aj, 158, 114
\bibitem[Han et al.(2001)]{Han2001} Han, C., Chang, H.-Y., An, J. H., \& Chang, K. 2001, \mnras, 328, 986
\bibitem[Han \& Gould(1995)]{Han1995}  Han, C., \& Gould, A.\ 1995, \apj, 447, 53
\bibitem[Han \& Gould(2003)]{Han2003}  Han, C., \& Gould, A.\ 2003, \apj, 592, 172
\bibitem[Han \& Jeong (1998)]{Han1998} Han, C., \& Jeong, Y. 1998, \mnras, 301, 231
\bibitem[Han et al.(2020)]{Han2020} Han, C., Kim, D., Jung, Y. K., et al. 2020, \aj, 160, 17
\bibitem[Han et al.(2022a)]{Han2022a} Han, C., Kim, D., Yang, H., et al. 2022a, \aap, submitted
\bibitem[Han et al.(2022b)]{Han2022b} Han, C., Udalski, A. Lee, C.-U., et al. 2022b, \aap, 658, A93
\bibitem[Han et al.(2013)]{Han2013} Han, C., Udalski, A., Choi, J.-Y., et al. 2013, \apj, 762, L28
\bibitem[Ida \& Lin(2010)]{Ida2010} Ida, S., \& Lin, D. N. C. 2010, \apj, 719, 810
\bibitem[Jung et al.(2021)]{Jung2021} Jung, Y.~K., Han, C., Udalski, A., et al.\ 2021, \aj, 161, 293
\bibitem[Jung et al.(2018)]{Jung2018} Jung, Y.~K., Udalski, A., Gould, A., et al.\ 2018, \aj, 155, 219
\bibitem[Kervella et al.(2004)]{Kervella2004} Kervella, P., Th\'evenin, F., Di Folco, E., \& S\'egransan, D.\ 2004, \aap, 426, 29
\bibitem[Kim et al.(2018)]{Kim2018} Kim, D.-J., Hwang, K.-H., Shvartzvald, et al. 2018, arXiv:1806.07545
\bibitem[Kim et al.(2016)]{Kim2016} Kim, S.-L., Lee, C.-U., Park, B.-G., et al.\ 2016, JKAS, 49, 37
\bibitem[Madsen \& Zhu(2019)]{Madsen2019} Madsen, S., \& Zhu, W. 2019, \apj, 878, L29
\bibitem[Mao \& Paczy\'nski(1991)]{Mao1991} Mao, S., \& Paczy\'nski, B. 1991, \apj, 374, L37
\bibitem[Mr\'oz et al.(2018)]{Mroz2018} Mr\'oz, Przemek, Ryu, Y.-H., Skowron, J., et al.\ 2018, \aj, 155, 121
\bibitem[Nataf et al.(2013)]{Nataf2013} Nataf, D.~M., Gould, A., Fouqu\'e, P., et al.\ 2013, \apj, 769, 88
\bibitem[Robin et al.(2003)]{Robin2003} Robin, A.~C., Reyl\'e, C., Derri\'ere, S., \& Picaud, S.\ 2003, \aap, 409, 523
\bibitem[Ryu et al.(2021)]{Ryu2021} Ryu, Y.-H., Mr\'oz, P., Gould, A., et al. 2021, \aj, 161, 126
\bibitem[Shvartzvald et al.(2017)]{Shvartzvald2017} Shvartzvald, Y., Yee, J. C., Calchi Novati, S., et al.\ 2017, \apj, 840, L3
\bibitem[Szyma\'nski et al.(2011)]{Szymanski2011} Szyma\'nski, M.K., Udalski, A., Soszy\'nski, I., et al. 2011, Acta Astron., 61, 83
\bibitem[Suzuki et al.(2018)]{Suzuki2018} Suzuki, D., Bennett, D. P., Udalski, A., et al. 2018, \aj, 155, 263
\bibitem[Tomaney \& Crotts(1996)]{Tomaney1996} Tomaney, A. B., \& Crotts, A. P. S.\ 1996, \aj, 112, 2872
\bibitem[Yee et al.(2012)]{Yee2012} Yee, J. C., Shvartzvald, Y., Gal-Yam, A., et al.\ 2012, \apj, 755, 102
\bibitem[Yoo et al.(2004)]{Yoo2004} Yoo, J., DePoy, D.~L., Gal-Yam, A., et al.\ 2004, \apj, 603, 139
\bibitem[Zang et al.(2021)]{Zang2021} Zang, W., Han, C., Kondo, I., et al. 2021, Research in Astronomy and Astrophysics, 21, 239
\bibitem[Zhu et al.(2014)]{Zhu2014} Zhu, W., Penny, M., Mao, S., Gould, A., \& Gendron, R. 2014, \apj, 788, 73
\vspace*{\fill}
\end{thebibliography}
\end{document}